\documentclass[epsfig,12pt]{article}
\usepackage{amsfonts}
\usepackage{amssymb}
\usepackage{graphicx}
\usepackage{amsmath}

\input epsf.sty
\newtheorem{theorem}{Theorem}
\newtheorem{acknowledgement}[theorem]{Acknowledgement}

\input{tcilatex}
\makeatletter \@addtoreset{equation}{section}
\renewcommand{\theequation}{\thesection.\arabic{equation}}
\begin{document}

\title{\textbf{Integrability and Generalized Monodromy Matrix }}
\author{T. Lhallabi$^{1,2,3}\thanks{%
lhallabi@fsr.ac.ma}$, A. Moujib$^{1,2,3}$ \\
{\small \textit{1.Laboratoire de Physique des Hautes Energies, Facul\'{e}
des Sciences, Rabat, Morocco,}}\\
{\small \textit{2.GNPHE, Groupement National de Physique des Hautes
Energies, }}\\
{\small \textit{Si\`{e}ge focal: Facul\'{e} des Sciences-Rabat, Morocco.}}\\
{\small \textit{3.Virtual African Centre for Basic Science and Technology,
VACBT, Rabat, Morocco}}}
\maketitle

\begin{abstract}
We construct the Generalized Monodromy matrix $\mathcal{\hat{M}}(\omega )$
of two dimensional string effective action by introducing the T-duality
group properties.The integrability conditions with general solutions
depending on spectral parameter are given. This construction is investigated
for the exactly solvable Wess, Zumino, Novikov and Witten (WZNW) model in
pp-wave Limit when B=0. \newline
\newline
\textbf{key words}: T-duality Symmetries, Two dimensional Non linear sigma
models, Integrability conditions, Monodromy Matrix.
\end{abstract}

\newpage

\section{ Introduction}

Two dimensional field theories play an important role in describing a
variety of physical systems \cite{1}-\cite{3}. Some of these models possess
the interesting property of integrability which corresponds to exactly
solvable models \cite{3}. The non linear $\sigma $-models in two dimensions
have been studied in great details from different perspectives \cite{2,3}.
Such models are endowed with a rich symmetry structure. However, the
symmetry structure of string theories is one of its most fascinating
features. In particular, duality symmetries have played a very important
role in the understanding of string dynamics \cite{4}-\cite{6}. The tree
level string effective action compactified on a $d$-dimensional torus $T$ $%
^{d}$ is known to be invariant under the non-compact global symmetry group $%
O(d,d).$ Furthermore, two dimensional models derived from dimensional
reduction of higher dimensional Einstein gravity as well as supergravity
theories have been studied to bring out their integrability properties \cite%
{7}-\cite{9}. In this context, one may recall that the construction of the
monodromy matrix turns out to be one of the principal objectives in the
study of integrable systems \cite{10,11}. The integrability of dimensionally
reduced gravity and supergravity to two dimensions have been studied
extensively by introducing the spectral parameter and construct a set of
currents which are invariant under a local $O(d)\times O(d)$ transformations
and satisfy the curvaturelessness condition \cite{6}.

The purpose of this article is to construct the generalized monodromy matrix 
$\mathcal{\hat{M}}(\omega )$ of a two-dimensional string effective action
obtained from a $D$-dimensional effective action, which is compactified on $%
T^{d}$, by considering the general integrability conditions.We provide the
transformation property of $\mathcal{\hat{M}}(\omega )$ under $O(d,d)$
group. In this process, we can synthesize the general classical
integrability properties and the $T$-duality symmetry of string theory in
order to derive a generalized monodromy matrix. The special properties of
generalized monodromy matrix are given in terms of general functions
depending on spectral parameter. Furthermore, the case of WZNW model with
pp-wave limit is studied where $\mathcal{\hat{M}}(\omega )$ acquiers a
special characteristic.

The paper is organized as follows: In section II, we recall the form of the
two-dimensional string effective action obtained by compactification on $%
T^{d}.$ Then,we consider a matricial form of background fields with $O(d,d)$
transformations which leave the dimensionally reduced action invariant
globally under $O(d,d)$ group and locally under $O(d)\times O(d)$. We devote
section III to the construction of the generalized monodromy matrix from
general integrability conditions in terms of general functions depending on
spectral parameter. Next, we give the transformations rules of $\mathcal{%
\hat{M}}(\omega )$ under $T$-duality symmetry and we present its explicit
form for simple background configurations.

In section IV, we investigate the generalized monodromy matrix in the WZNW
model with pp-wave limit. The summary and conclusion are given in section V.

\section{Two-Dimensional String Effective Action}

We are interested by the two-dimensional $\sigma $-model coupled to gravity
namely%
\begin{equation}
S=\int dx^{2}\sqrt{-g}e^{-\bar{\Phi}}\left[ R+(\partial \bar{\Phi})^{2}+%
\frac{1}{8}Tr(\partial _{\alpha }M^{-1}\partial ^{\alpha }M)\right] 
\label{r}
\end{equation}%
which is derived by dimensional reduction of the effective action of string
theory \cite{12,13}. In this expression, $g$ is the determinant of the
metric $g_{\alpha \beta }$ where $\alpha ,\beta =0,1$ are the two-
dimensional space time indices and $R$ is the associated Ricci scalar
curvature. The field $\bar{\Phi}$ is the usual shifted dilaton given by%
\begin{equation}
\bar{\Phi}=\Phi -\frac{1}{2}\log \det G_{ij}
\end{equation}%
where $G_{ij}$ is the metric in the internal space, corresponding to the
toroidally compactified coordinates $x^{i}$ ; $i,j=2,3,...(D-1)$ with $d=D-2$%
. $M$ is a $2d\times 2d$ symmetric matrix of the form%
\begin{equation}
M=\left( 
\begin{array}{cc}
G^{-1} & -G^{-1}B \\ 
BG^{-1} & G-BG^{-1}B%
\end{array}%
\right) 
\end{equation}%
where $B_{ij}$ is the moduli coming from dimensional reduction of the Neveu
Schwartz-Neveu Schwartz sector in higher space time dimensions. $G$ and $B$
parametrize the coset $O(d,d)/O(d)\times O(d)$ and the matrix $M$
corresponds to a symmetric representation of the group $O(d,d)$.\newline
The action (\ref{r}) is invariant under global $O(d,d)$ transformations
namely 
\begin{equation*}
g_{\alpha \beta }\longrightarrow g_{\alpha \beta }
\end{equation*}%
\begin{equation}
\bar{\Phi}\longrightarrow \bar{\Phi}
\end{equation}%
\begin{equation*}
M\longrightarrow \Omega ^{T}M\Omega \ \ \quad \ ,\quad \ \ \Omega \in O(d,d)
\end{equation*}%
and the variation with respect to $M$ leads to the conservation law 
\begin{equation}
\partial _{\alpha }\left[ e^{-\bar{\Phi}}\sqrt{-g}g^{\alpha \beta
}M^{-1}\partial _{\beta }M\right] =0  \label{5}
\end{equation}%
In order to make apparent a local $O(d)\times O(d)$ as well as a global $%
O(d,d)$ transformations, it is convenient to introduce a triangular matrix $V
$ contained in $O(d,d)/O(d)\times O(d)$ of the following form%
\begin{equation}
V=\left( 
\begin{array}{cc}
E^{-1} & 0 \\ 
BE^{-1} & E^{T}%
\end{array}%
\right) 
\end{equation}%
such that $M=VV^{T}$ with $(E^{T}E)_{ij}=G_{ij}$ where $E$ is the vielbein
in the internal space. Consequently, the matrix $V$ parametrizing the coset $%
O(d,d)/O(d)\times O(d)$ transforms non trivially under global $O(d,d)$ and
local $O(d)\times O(d)$ namely%
\begin{equation}
V\longrightarrow \Omega ^{T}Vh(x)
\end{equation}%
where $\Omega \in O(d,d)$ and $h(x)\in O(d)\times O(d)$.\newline
Let us remark that the matrix $M$ is sensitive only to a global $O(d,d)$
rotation. Furthermore, the matter sector of this model governed by the third
term in (\ref{r}) is based on a set of scalar fields which are combined into
a matrix $V(x)$ taking values in a non-compact Lie group $G$ with the
maximal compact subgroup $H$. This subgroup can be characterized by means of
a symmetric space involution $\eta :$ \ \ $G\longrightarrow G$ \ as follows%
\begin{equation}
H=\{h\in G/\ \eta (h)=h\}
\end{equation}%
naturally, the involution extends to the corresponding Lie algebras $g=LieG$
and $h=LieH$ where $G=O(d,d)$ and its maximal compact subgroup $H=O(d)\times
O(d)$ characterized by the involution \cite{10}%
\begin{equation}
\eta (g)\equiv (g^{T})^{-1}
\end{equation}%
for $g\in G$.\newline
However, from the matrix $V$, we can construct the following current \cite%
{15,16} 
\begin{equation}
V^{-1}\partial _{\alpha }V=P_{\alpha }+Q_{\alpha }  \label{4}
\end{equation}%
which belongs to the Lie algebra of $O(d,d)$. In such decomposition $%
Q_{\alpha }$ belongs to the Lie algebra of the maximally compact subgroup $%
O(d)\times O(d)$ and $P_{\alpha }$ to the complement. From the symmetric
space automorphism property of the coset $O(d,d)/O(d)\times O(d)$ it follows
that 
\begin{equation}
P_{\alpha }^{T}=P_{\alpha }
\end{equation}%
\ 
\begin{equation*}
Q_{\alpha }^{T}=-Q_{\alpha }
\end{equation*}%
Therefore 
\begin{equation}
P_{\alpha }=\frac{1}{2}\left[ V^{-1}\partial _{\alpha }V+(V^{-1}\partial
_{\alpha }V)^{T}\right]   \label{2}
\end{equation}%
\begin{equation*}
Q_{\alpha }=\frac{1}{2}\left[ V^{-1}\partial _{\alpha }V-(V^{-1}\partial
_{\alpha }V)^{T}\right] 
\end{equation*}%
Now it is easy to show that 
\begin{equation}
Tr\left[ \partial _{\alpha }M^{-1}\partial _{\beta }M\right] =-4Tr\left[
P_{\alpha }P_{\beta }\right]   \label{3}
\end{equation}%
The currents in (\ref{2}) and the action (\ref{r}) are invariant under local
gauge transformation%
\begin{equation}
V\longrightarrow Vh(x)
\end{equation}%
where $h(x)\in O(d)\times O(d)$. Consequently, the composite fields $%
P_{\alpha }$ and $Q_{\alpha }$ which are inert under rigid $O(d,d)$
invariance transform under $O(d)\times O(d)$ according to \cite{16}.%
\begin{eqnarray}
P_{\alpha } &\longrightarrow &h^{-1}(x)P_{\alpha }h(x) \\
Q_{\alpha } &\longrightarrow &h^{-1}(x)Q_{\alpha }h(x)+h^{-1}(x)\partial
_{\alpha }h(x)  \notag
\end{eqnarray}%
We note that $Q_{\alpha }$ transforms like a gauge field while $P_{\alpha }$
transforms as belonging to the adjoint representation. It is clear that (\ref%
{3}) is invariant under the global $O(d,d)$ as well as the local $O(d)\times
O(d)$ transformations. This fact allows to obtain the integrability
conditions and the derivation of the general monodromy matrix which will be
the subject of the next section.

\section{Classical Integrability and General Monodromy Matrix}

In this section we consider the case of two dimensional sigma model in flat
space time defined on the coset $O(d,d)/O(d)\times O(d)$. The integrability
conditions following from the currents (\ref{4}) corresponds to the zero
curvature condition namely 
\begin{equation}
\partial _{\alpha }\left[ V^{-1}\partial _{\beta }V\right] -\partial _{\beta
}\left[ V^{-1}\partial _{\alpha }V\right] +\left[ (V^{-1}\partial _{\alpha
}V),(V^{-1}\partial _{\beta }V)\right] =0  \label{6}
\end{equation}%
Such equation with the definitions (\ref{2}) of $P_{\alpha }$ and $Q_{\alpha
}$ are subject to the compatibility relations 
\begin{equation}
\partial _{\alpha }Q_{\beta }-\partial _{\beta }Q_{\alpha }+\left[ Q_{\alpha
},Q_{\beta }\right] =-\left[ P_{\alpha },P_{\beta }\right]  \label{7}
\end{equation}%
\begin{equation}
D_{\alpha }P_{\beta }-D_{\beta }P_{\alpha }=0
\end{equation}%
with%
\begin{equation}
D_{\alpha }P_{\beta }=\partial _{\alpha }P_{\beta }+\left[ Q_{\alpha
},P_{\beta }\right]
\end{equation}%
For the case of flat space time, the equation of motion (\ref{5}) modifies to

\begin{equation}
\eta ^{\alpha \beta }\partial _{\alpha }\left[ M^{-1}\partial _{\beta }M%
\right] =0
\end{equation}%
which is equivalent to%
\begin{equation}
\eta ^{\alpha \beta }D_{\alpha }P_{\beta }=0
\end{equation}%
In the same way as in ref \cite{14}-\cite{16}, we introduce a one parameter
family of matrices with a constant spectral parameter $t$ such that%
\begin{equation}
\hat{V}(x,t=0)=\hat{V}(x)
\end{equation}%
Now, let us consider the generalized current decomposition with arbitrary
functions of the spectral parameter $t$ as follows%
\begin{equation}
\hat{V}^{-1}\partial _{\alpha }\hat{V}=Q_{\alpha }+f(t)P_{\alpha
}+g(t)\varepsilon _{\alpha \beta }P^{\beta }
\end{equation}%
where $f(t)$ and $g(t)$ are general functions satisfying the following
conditions%
\begin{eqnarray}
f(t &=&0)=1\ \ \ \ \ \ \ \quad \ ,\ \ \ \quad \quad \underset{%
t\longrightarrow +\infty }{\lim }f(t)=-1  \label{10} \\
g(t &=&0)=0\ \ \ \ \ \quad \ ,\ \ \ \ \ \ \ \quad \underset{t\longrightarrow
+\infty }{\lim }g(t)=0
\end{eqnarray}%
in flat space-time. Let us note that these functions have to possess
singularities for $t=\pm 1$ in order to recover the case of ref \cite{16}.
Then, the integrability conditions (\ref{6}) are rewritten as%
\begin{equation}
\partial _{\alpha }\left[ \hat{V}^{-1}\partial _{\beta }\hat{V}\right]
-\partial _{\beta }\left[ \hat{V}^{-1}\partial _{\alpha }\hat{V}\right] +%
\left[ (\hat{V}^{-1}\partial _{\alpha }\hat{V}),(\hat{V}^{-1}\partial
_{\beta }\hat{V})\right] =0
\end{equation}%
which can be expressed in terms of the functions $f$ and $g$ as follows%
\begin{equation}
\partial _{\alpha }Q_{\beta }-\partial _{\beta }Q_{\alpha }+2\left[
Q_{\alpha },Q_{\beta }\right] +\partial _{\alpha }\left[ f(t)+g(t)\right]
P_{\beta }-\partial _{\beta }\left[ f(t)+g(t)\right] P_{\alpha }
\end{equation}%
\begin{equation*}
+\left[ f(t)+g(t)\right] \left( \partial _{\beta }P_{\alpha }-\partial
_{\alpha }P_{\beta }+2\left[ P_{\alpha },P_{\beta }\right] \right) +4f(t)g(t)%
\left[ P_{\alpha },P_{\beta }\right] =0
\end{equation*}%
By using the relation (\ref{7}), such equations take the following form.%
\begin{equation}
\begin{array}{c}
(f^{\prime }+g^{\prime })\left[ \partial _{\alpha }tP_{\beta }-\partial
_{\beta }tP_{\alpha }\right] +(f+g)\left( \partial _{\beta }P_{\alpha
}-\partial _{\alpha }P_{\beta }+2\left[ P_{\alpha },P_{\beta }\right]
\right)  \\ 
+\left( 4fg-1\right) \left[ P_{\alpha },P_{\beta }\right] +\left[ Q_{\alpha
},Q_{\beta }\right] =0%
\end{array}%
\end{equation}%
This leads to the integrability condition of the spectral parameter namely%
\begin{equation}
(\partial _{\beta }t)P_{\alpha }-(\partial _{\alpha }t)P_{\beta }=(f\text{ }%
^{\prime }+g^{\prime })^{-1}\left\{ 
\begin{array}{c}
(f+g)(\partial _{\beta }P_{\alpha }-\partial _{\alpha }P_{\beta }+2\left[
P_{\alpha },P_{\beta }\right] ) \\ 
+\left( 4fg-1\right) \left[ P_{\alpha },P_{\beta }\right] +\left[ Q_{\alpha
},Q_{\beta }\right] 
\end{array}%
\right\} 
\end{equation}%
which can be simplified by using the light cone indices to the following
equations%
\begin{equation}
\partial _{\pm }t=P_{0}^{-1}\left( P_{\pm }\pm H(t)\right) 
\end{equation}%
with%
\begin{equation}
H(t)=\left( f^{\prime }\text{ }+g^{\prime }\right) ^{-1}\left\{ 
\begin{array}{c}
(f+g)(\partial _{1}P_{0}-\partial _{0}P_{1}+2\left[ P_{0},P_{1}\right] )+ \\ 
\left( 4fg-1\right) \left[ P_{0},P_{1}\right] +\left[ Q_{0},Q_{1}\right] 
\end{array}%
\right\} 
\end{equation}%
Furthermore, in order to obtain the monodromy matrix in terms of generalized
functions we can parametrize the $P_{\pm }$ quantities in terms of the
vielbein as follows%
\begin{equation}
P_{\pm }=\left( 
\begin{array}{cc}
-E^{-1}\partial _{\pm }E & 0 \\ 
0 & E^{-1}\partial _{\pm }E%
\end{array}%
\right) \ \ \quad \ ,\quad \ \ Q_{\pm }=0
\end{equation}%
These currents that are subject of integrability condition can be
incorporated in the matrix $\hat{V}$ as%
\begin{equation}
\hat{V}^{-1}\partial _{\pm }\hat{V}=\left( f(t)+g(t)\right) \left( 
\begin{array}{cc}
-E^{-1} & 0 \\ 
0 & E^{-1}%
\end{array}%
\right) \partial _{\pm }E
\end{equation}%
Let us note that in this parametrization, the antisymmetric tensor field is
chosen to be zero and the matrices $E$ and $G$ are assumed to be diagonal
namely \cite{16}%
\begin{equation}
E=diag\left( e^{\frac{1}{2}(\lambda +\psi _{1})},e^{\frac{1}{2}(\lambda
+\psi _{2})},....,e^{\frac{1}{2}(\lambda +\psi _{d})}\right) 
\end{equation}%
\begin{equation*}
G=diag\left( e^{\lambda +\psi _{1}},e^{\lambda +\psi _{2}},....,e^{\lambda
+\psi _{d}}\right) 
\end{equation*}%
with $\sum \psi =0$ so that%
\begin{equation}
\lambda =\frac{1}{d}\log \det G
\end{equation}%
We shall assume that the scalar fields live on a finite dimensional (non
compact) symmetric space $G/H$ and that the local $H$ invariance has been
already restored. They are thus described by a matrix $V(x)$ subject to the
transformation \cite{14}.%
\begin{equation}
V(x)\longrightarrow KV(x)h(x)
\end{equation}%
with arbitrary $K\in O(d,d)$ and $h(x)\in O(d)\times O(d)$. The dynamics is
defined via the Lie algebra decomposition (\ref{4}) and in addition to the
matter fields, the model contains a dilaton field $\rho $ which is given in
the conformal gauge by%
\begin{equation}
\rho (x)=\rho _{+}(x^{+})+\rho _{-}(x^{-})=e^{-\bar{\Phi}}
\end{equation}%
with%
\begin{equation*}
\rho =\det E
\end{equation*}%
where $\rho _{+}(x^{+})$ and $\rho _{-}(x^{-})$ are left and right moving
solutions. The matter field equations of motion read%
\begin{equation}
D^{\alpha }(\rho P_{\alpha })=0  \label{8}
\end{equation}%
where $D_{\alpha }$ is the $O(d)\times O(d)$ covariant derivative. There are
two first order equations namely%
\begin{equation}
\partial _{+}\rho \partial _{+}\sigma =\frac{1}{2}\rho Tr\left(
P_{+}P_{+}\right) 
\end{equation}%
\begin{equation*}
\partial _{-}\rho \partial _{-}\sigma =\frac{1}{2}\rho Tr\left(
P_{-}P_{-}\right) 
\end{equation*}%
where $\sigma $ is defined by%
\begin{equation}
\sigma \equiv \log \lambda -\frac{1}{2}\log \left( \partial _{+}\rho
\partial _{-}\rho \right) 
\end{equation}%
Furthermore, the matter field equation (\ref{8}) for $\hat{V}(x,t)$ as well
as the usual integrability conditions following from (\ref{4}) can be
recovered as the compatibility conditions of the linear system (Lax pair) 
\cite{14}.%
\begin{equation}
\hat{V}^{-1}D_{\alpha }\hat{V}=\left( f(t)+g(t)\right) P_{\alpha }  \label{9}
\end{equation}%
\begin{equation*}
Q_{\alpha }=0
\end{equation*}%
where%
\begin{equation}
D_{\alpha }\hat{V}=\partial _{\alpha }\hat{V}-\hat{V}Q_{\alpha }
\end{equation}%
The consistency of (\ref{9}) requires that the spectral parameter $t$ itself
be subject to a very similar system of linear differential equations as
follows%
\begin{equation}
t^{-1}\partial _{\pm }t=\left( f(t)+g(t)\right) \rho ^{-1}\partial _{\pm
}\rho 
\end{equation}%
that relates $\rho $ and $t$ as a B\"{a}cklund duality \cite{14}. Indeed%
\begin{equation}
\partial _{\pm }t=t\left( f(t)+g(t)\right) e^{\bar{\Phi}}\partial _{\pm }e^{-%
\bar{\Phi}}
\end{equation}%
Thus, from the current form, we assume that%
\begin{equation}
f(t)+g(t)=\sqrt{\frac{\omega -\rho _{-}}{\omega +\rho _{+}}}
\end{equation}%
where $\omega $ is a function of spectral parameter and represent the
constant of integration of $t$ ($\omega $ is namely as a hidden spectral
parameter) given by%
\begin{equation}
\omega =-\rho _{+}+\frac{e^{-\bar{\Phi}}}{1-N^{2}(t)}
\end{equation}%
where $N(t)$ must verify that $N(t)\neq \pm 1$ with $N(t)=f(t)+g(t)$.
Therefore, the generalized monodromy matrix has to be of the form%
\begin{equation}
\mathcal{\hat{M}}=\hat{V}(x,t)\hat{V}^{T}(x,\frac{1}{t})=\left( 
\begin{array}{cc}
\mathcal{M}(\omega ) & 0 \\ 
0 & \mathcal{M}^{-1}(\omega )%
\end{array}%
\right) 
\end{equation}%
where $\mathcal{M}(\omega )$ is diagonal with%
\begin{equation}
\mathcal{M}_{i}(\omega )=\frac{\omega _{i}-\omega }{\omega _{i}+\omega }
\end{equation}%
and%
\begin{equation}
\omega _{i}=-\rho _{+}+\frac{e^{-\bar{\Phi}}}{1-N^{2}(t_{i})}
\end{equation}%
Finally this generalized monodromy matrix can be used to examine general
currents and general integrability conditions in some cosmological models.
In fact, in the next section we study the WZNW model in the pp-wave limit
where the generalized monodromy matrix has an interesting behaviour.

\section{Generalized Monodromy matrix in pp-wave limit}

The WZNW model \cite{17,18} is one of the convenient models for constructing
the generalized monodromy matrix in pp-wave limit. In fact, by investigating
the previous results we can start with%
\begin{equation}
\Omega =\left( 
\begin{array}{cccc}
\sqrt{1+q_{0}} & 0 & 0 & -\sqrt{q_{0}} \\ 
0 & \sqrt{1+q_{0}} & \sqrt{q_{0}} & 0 \\ 
0 & \sqrt{q_{0}} & \sqrt{1+q_{0}} & 0 \\ 
-\sqrt{q_{0}} & 0 & 0 & \sqrt{1+q_{0}}%
\end{array}%
\right) 
\end{equation}%
which is belonging to $O(d,d)$ group and where $q_{0}$ is a constant
parameter parametrizing the embedding of $U(1)$ into $E_{2}^{c}\otimes U(1)$ 
\cite{18,19}. $E_{2}^{c}$ represents the two- dimensional Euclidean group
with a central extension.\newline
As we have seen in the previous section, the one parameter family of
potentials $\hat{V}(x,t)$ for $B=0$ satisfy%
\begin{equation}
\hat{V}^{-1}\partial _{\pm }\hat{V}=(f(t)+g(t))P_{\pm }
\end{equation}%
and its diagonal form can be written as%
\begin{equation}
\hat{V}(x,t)=diag(\hat{V}_{1},\hat{V}_{2},\hat{V}_{3},\hat{V}_{4})
\end{equation}%
since $\hat{V}(x,t)$ transforms globally under $O(d,d)$ transformations
through $\Omega $ matrix and locally under $O(d)\times O(d)$ transformations
by $h(x)$. This latter is an element of the maximal compact subgroup $H$ of
the T-duality group \cite{19}. Hence we can write%
\begin{equation}
\hat{V}(x,t)=\Omega ^{T}\hat{V}^{(B=0)}h(x)=\left( 
\begin{array}{cccc}
U_{1} & 0 & 0 & U_{2} \\ 
0 & U_{3} & U_{4} & 0 \\ 
0 & U_{5} & U_{6} & 0 \\ 
U_{7} & 0 & 0 & U_{8}%
\end{array}%
\right) 
\end{equation}%
where $h(x)$ is given in the Novikov-Witten model \cite{19} by%
\begin{equation}
h(x)=\left( 
\begin{array}{cccc}
\cos \theta  & 0 & 0 & \sin \theta  \\ 
0 & \cos \theta  & -\sin \theta  & 0 \\ 
0 & \sin \theta  & \cos \theta  & 0 \\ 
-\sin \theta  & 0 & 0 & \cos \theta 
\end{array}%
\right) 
\end{equation}%
This allows to obtain the matrix elements of $\hat{V}(x,t)$ namely%
\begin{equation*}
U_{1}=\sqrt{1+q_{0}}\cos \theta +\sqrt{q_{0}}\hat{V}_{4}\sin \theta 
\end{equation*}%
\begin{equation*}
U_{2}=\sqrt{1+q_{0}}\sin \theta -\sqrt{q_{0}}\hat{V}_{4}\cos \theta 
\end{equation*}%
\begin{equation}
U_{3}=\sqrt{q_{0}}\sin \theta +\sqrt{1+q_{0}}\hat{V}_{2}\cos \theta 
\end{equation}%
\begin{equation*}
U_{4}=\sqrt{q_{0}}\cos \theta -\sqrt{1+q_{0}}\hat{V}_{2}\sin \theta 
\end{equation*}%
\begin{equation*}
U_{5}=\sqrt{1+q_{0}}\sin \theta +\sqrt{q_{0}}\hat{V}_{2}\cos \theta 
\end{equation*}%
\begin{equation*}
U_{6}=\sqrt{1+q_{0}}\cos \theta -\sqrt{q_{0}}\hat{V}_{2}\sin \theta 
\end{equation*}%
\begin{equation*}
U_{7}=-\sqrt{q_{0}}\cos \theta +\sqrt{1+q_{0}}\hat{V}_{4}\sin \theta 
\end{equation*}%
\begin{equation*}
U_{8}=-\sqrt{q_{0}}\sin \theta +\sqrt{1+q_{0}}\hat{V}_{4}\cos \theta 
\end{equation*}%
Consequently, this leads to the generalized monodromy matrix of the form%
\begin{equation}
\mathcal{\hat{M}}^{(B)}=\hat{V}^{(B)}(x,t)(\hat{V}^{(B)})^{T}(x,\frac{1}{t}%
)=\left( 
\begin{array}{cccc}
C_{1} & 0 & 0 & C_{2} \\ 
0 & C_{3} & C_{4} & 0 \\ 
0 & C_{4} & C_{5} & 0 \\ 
C_{2} & 0 & 0 & C_{6}%
\end{array}%
\right) 
\end{equation}%
where the coefficients are obtained after some compute by%
\begin{eqnarray}
C_{1} &=&1+\frac{2q_{0}}{1-2\omega }  \notag \\
C_{2} &=&-\sqrt{q_{0}(1+q_{0})}-\sqrt{q_{0}}\frac{1+2\omega }{1-2\omega } 
\notag \\
C_{3} &=&-1+\frac{2(1+q_{0})}{1+2\omega } \\
C_{4} &=&\sqrt{q_{0}}+\sqrt{q_{0}(1+q_{0})}\frac{1-2\omega }{1+2\omega } 
\notag \\
C_{5} &=&1+\frac{2q_{0}}{1+2\omega }  \notag \\
C_{6} &=&-1+\frac{2(1+q_{0})}{1-2\omega }  \notag
\end{eqnarray}%
with%
\begin{equation*}
\omega =-\rho _{+}+\frac{e^{-\bar{\Phi}}}{1-N^{2}(t)}
\end{equation*}%
We remark that the generalized monodromy matrix for the particular case $B=0$
and when $q_{0}=0$ is given by%
\begin{equation}
\mathcal{\hat{M}(}\omega )=\left( 
\begin{array}{cccc}
1 & 0 & 0 & 0 \\ 
0 & \frac{1-2\omega }{1+2\omega } & 0 & 0 \\ 
0 & 0 & 1 & 0 \\ 
0 & 0 & 0 & \frac{1+2\omega }{1-2\omega }%
\end{array}%
\right) 
\end{equation}%
However, the generalized monodromy matrix in the pp-wave limit for this
model can be constructed by following the same procedure as before but by
taking $x^{-}\longrightarrow 0$ as limit in consideration. In fact, the
background consist of the metric with a vanishing $B$ field, of the form 
\cite{9}%
\begin{equation}
dS^{2}=-2dx^{+}dx^{-}+\tan ^{-2}\frac{x^{+}}{\sqrt{2}}dy^{2}+\tan ^{2}\frac{%
x^{-}}{\sqrt{2}}dz^{2}
\end{equation}%
Therefore, the diagonal form of the $\hat{V}$ matrix can be given as follows 
\cite{19}%
\begin{equation}
\hat{V}^{(B=0)}(x,t)=diag(\hat{V}_{1},\hat{V}_{2},\hat{V}_{3},\hat{V}%
_{4})=diag(1,\frac{(1-t)\tan x^{+}}{t+\tan ^{2}x^{+}},1,\frac{t+\tan
^{2}x^{+}}{(1-t)\tan x^{+}})
\end{equation}%
Then, in the pp-wave limit $x^{-}\longrightarrow 0$, where $x=\frac{x^{+}}{%
\sqrt{2}}$ we deduce that%
\begin{equation}
\hat{V}_{1}=-\frac{t-1}{t\tan ^{2}\frac{x^{+}}{\sqrt{2}}+1}=\hat{V}_{3}^{-1}
\end{equation}%
\begin{equation}
\hat{V}_{2}=-\frac{(t-1)\tan ^{2}\frac{x^{+}}{\sqrt{2}}}{t-\tan ^{2}\frac{%
x^{+}}{\sqrt{2}}}=\hat{V}_{4}^{-1}
\end{equation}%
In particular for $t=0$, this return to $\hat{V}-$matrix which is appeared
before. It follows that the generalized monodromy matrix is now determined by%
\begin{equation}
\mathcal{\hat{M}}^{(NW)}\mathcal{(}\omega )=\hat{V}(x,t)\hat{V}^{T}(x,\frac{1%
}{t})=\left( 
\begin{array}{cccc}
\frac{1-2\omega }{1+2\omega } & 0 & 0 & 0 \\ 
0 & \frac{1-2\omega }{1+2\omega } & 0 & 0 \\ 
0 & 0 & \frac{1+2\omega }{1-2\omega } & 0 \\ 
0 & 0 & 0 & \frac{1+2\omega }{1-2\omega }%
\end{array}%
\right) 
\end{equation}%
where $\omega $ is given in the pp-wave limit by%
\begin{equation}
\omega =-\rho _{+}(x^{+})+\frac{e^{-\bar{\Phi}}}{1-N^{2}(t)}
\end{equation}%
with%
\begin{equation}
N(t)=f(t)+g(t)\ \ \ \ \quad \ ,\ \ \quad \ N(t)\neq \pm 1
\end{equation}%
In this case, $\omega $ depends only on one light-cone coordinate. and%
\begin{equation}
e^{-\bar{\Phi}}=\rho (\frac{x^{+}}{\sqrt{2}})=\rho _{+}(x^{+})+\rho _{-}(0)
\end{equation}%
Hence, if we choose $\rho _{-}(0)=0$ we have%
\begin{equation}
e^{-\bar{\Phi}}=\rho (\frac{x^{+}}{\sqrt{2}})=\rho _{+}(x^{+})
\end{equation}%
which means that the dilaton is expressed with only one light-cone
coordinate and admits just one part of solution. We note that the
generalized monodromy matrix in the pp-wave limit which depends only of one
light cone coordinate $x^{+}$ still admits degenerate poles in the NW model.

\section{ Conclusion}

In this paper, we have constructed the generalized monodromy matrix of
two-dimensional string effective action by using the general integrability
conditions which are expressed in terms of general functions $f$ and $g$
depending on the spectral parameter and satisfying conditions (\ref{10}) in
the flat space. One of our principal aims is to take into account the
symmetries associated with the string effective action for constructing the
generalized monodromy matrix which contains information about these
symmetries. In fact, we have shown that the generalized monodromy matrix
transforms non-trivially under the non-compact $T$-duality group and possess
a special structure in pp-wave limit for the WZWN model. Furthermore, it is
necessary to note that we have realized a connection between $T$-duality
symmetry and general integrability properties by introducing a spectral
parameter $t$ which is space-time dependent. Indeed, we have envisaged a
vielbein $E(x,t)$ in order to define the action under consideration and
examine its invariance globally under $G$ and locally under $H$. By
including $E$, we have introduced an infinite potentials family of matrices
parametrized by general functions depending on a continuous spectral
parameter necessary to obtain the generalized monodromy matrix in ordinary
and pp-wave limit cases. Finally, our general results have likely
exploitations and more applications in string theory and cosmological models 
\cite{20}-\cite{23} such the Becchi-Rouet-Stora-Tyutin quantization of
bosonic string on $AdS_{5}\times S^{5}$ and analysis of black-holes with
solutions. Furthermore the integrable models like calogero model for two
particles and $N$ particles and others cosmological models with non-compact
Lie groups can be studied. With these results, a very large and rich class
of models can be treated from this process.

\begin{acknowledgement}
We would like to thank the program Protars III D12/25, CNRST for it's
contribution to this research.
\end{acknowledgement}

\end{document}